# A Simple Multi-Directional Absorbing Layer Method to Simulate Elastic Wave Propagation in Unbounded Domains

*Jean-François Semblat, Luca Lenti, Ali Gandomzadeh.*

*Université Paris-Est, Laboratoire Central des Ponts et Chaussées, 58 bd Lefebvre, 75015 Paris, France (semblat@lcpc.fr)*

**Abstract:** The numerical analysis of elastic wave propagation in unbounded media may be difficult due to spurious waves reflected at the model artificial boundaries. This point is critical for the analysis of wave propagation in heterogeneous or layered solids. Various techniques such as Absorbing Boundary Conditions, infinite elements or Absorbing Boundary Layers (e.g. Perfectly Matched Layers) lead to an important reduction of such spurious reflections.

In this paper, a simple absorbing layer method is proposed: it is based on a Rayleigh/Caughey damping formulation which is often already available in existing Finite Element softwares. The principle of the *Caughey Absorbing Layer Method* is first presented (including a rheological interpretation). The efficiency of the method is then shown through 1D Finite Element simulations considering homogeneous and heterogeneous damping in the absorbing layer. 2D models are considered afterwards to assess the efficiency of the absorbing layer method for various wave types and incidences. A comparison with the PML method is first performed for pure P-waves and the method is shown to be reliable in a more complex 2D case involving various wave types and incidences. It may thus be used for various types of problems involving elastic waves (e.g. machine vibrations, seismic waves, etc).

**Nomenclature:**

| | |
|---|---|
| $a_0$, $a_1$ | Rayleigh coefficients |
| $[C]$ | damping matrix |
| $E$ | Young modulus |
| $f$ | frequency |
| $f_R$ | fundamental frequency (Ricker wavelet) |
| $[K]$ | stiffness matrix |
| **k** | wave vector |
| $[M]$ | mass matrix |
| $M$ | complex modulus |
| $Q$ | quality factor |
| $t$ | time |
| $t_s$ | time shift (Ricker wavelet) |
| $t_p$ | fundamental period (Ricker wavelet) |
| **u** | displacement vector |
| $\{u\}$ | vector of nodal displacements |
| $\{\dot{u}\}$ | vector of nodal velocities |
| $\{\ddot{u}\}$ | vector of nodal accelerations |
| **x** | position vector |
| $\alpha$ | attenuation vector |
| $\zeta$ | viscosity |
| $\eta$ | loss factor |
| $\lambda$ | wavelength |
| $\xi$ | damping ratio |
| $\omega$ | circular frequency |

## 1. Modelling wave propagation in unbounded domains

### 1.1 Numerical methods for elastic waves

Various numerical methods are available to simulate elastic wave propagation in solids: finite differences [1,2], finite elements [3,4], boundary elements [5,6], spectral elements [7,8], meshfree methods [9], etc. Such methods as finite or spectral elements have strong advantages (for complex geometries, nonlinear media, etc) but also important drawbacks such as numerical dispersion [10,11,12] for low order finite elements, or spurious reflections at the mesh boundaries [4,13]. Other methods such as the Boundary Element Method (BEM) are generally limited to weakly heterogeneous linear media but allow an accurate description of the radiated waves at infinity [6,14,15]. Another limitation of the classical BEM is that it leads to unsymmetric fully populated matrices. New efficient Fast BEM formulations have been recently proposed to model 3D elastic wave propagation [16,17].

### 1.2 Methods to deal with spurious reflections

The problem of spurious reflections may be dealt with using the BEM or coupling it with another numerical method [18]. At very large scales, domain Reduction Methods are also available in the framework of Finite Element approaches [19].

Another alternative is to directly attenuate the spurious reflections at the mesh boundaries considering *Non Reflecting Boundary Conditions (NRBCs)*. A detailed review of such techniques was proposed by Givoli [20] and Harari and Shohet [21] and a recent paper by Festa and Vilotte [22] gives additional references. The various *NRBCs* approaches may be characterized the following way:

- *Absorbing boundary conditions* ("*ABCs*"): they involve specific conditions at the model boundaries to approximate the radiation condition for the elastic waves [4,13]. Engquist and Majda [23] proposed a technique based on local approximate boundary conditions of increasing order (also see the work of Bayliss and Turkel [24]). Since these techniques involve "one-way" boundary operators (i.e. along one direction only), it is difficult to avoid spurious reflections due to various wave types involving different polarizations. Furthermore, the extension of such techniques to elastic waves leads to complex systems of equations difficult to analyze in terms of stability [22]. Givoli [20] also proposed *NRBC*s (Dirichlet-to-Neumann, "*DtN*", operator) that are nonlocal in space or in time (or both) and may be considered as more effective than local boundary conditions (independent of the angle of incidence).
- *Infinite elements*: they allow the approximation of the decaying laws governing the waves radiation process at infinity [25]. The principle is to use finite elements with their end nodes moved at infinity. Their drawbacks are similar to that of the absorbing boundaries.
- *Absorbing boundary layers* ("*ABLs*"): such methods (e.g. Perfectly Matched Layers or *PMLs*) have been widely studied in the recent years [26,27,28,29,30]. The PML technique, introduced by Bérenger in the field of electromagnetics [31], is based on the description of attenuating properties along a specific direction in an absorbing layer of finite thickness located at the medium boundaries. Inside the PML, a field described by the plane wave $\exp[-i(\mathbf{k}\cdot\mathbf{x} - \omega t)]$ decreases, in the $x_i$ direction, by a factor generally independent of frequency [22]:

$$\exp\left(-\frac{k_i}{\omega}\int \alpha_i(\zeta)d\zeta\right) \quad (1)$$

  where $\alpha_i$ is an analytical function.
- The extension of PML techniques to acoustic and elastic wave propagation was proposed by various authors [27,32,33,34]. Classical PMLs are more efficient than "*ABCs*" but several cases lead to some instabilities: grazing incidence, shallow models involving surface waves, anisotropic media. Several improved PML formulations were recently proposed to overcome such difficulties. The filtering or convolutional PML allows the treatment of surface waves in shallow media [27]. In this formulation, the field decay in the $x_i$ direction is governed by a factor depending on frequency [22]:

$$\exp\left(-\frac{k_i}{\omega}\frac{\omega^2 - i\omega\omega_c}{\omega^2 + \omega_c^2}\int \alpha_i(\zeta)d\zeta\right) \quad (2)$$

where $\omega_c$ is a cut-off frequency.

The multi-directional PML formulation was recently proposed to deal with grazing incidences and strong anisotropies [30]. It allows various choices for the attenuation vector α in the absorbing layer:

$$\mathbf{u} = \mathbf{A}\exp(-\boldsymbol{\alpha}\cdot\mathbf{x})\exp[-i(\mathbf{k}\cdot\mathbf{x} - \omega t)] \quad (3)$$

where **u** is the displacement vector, **A** the polarization vector, **x** the position vector, **k** the wave vector, $\omega$ the frequency and $t$ the time.

It is thus possible to deal with various wave types (i.e. polarizations) and incidences. The multi-directional PML formulation leads to better efficiency and numerical stability.

In this paper, we propose a simple and reliable absorbing layer method to reduce the spurious reflections at the model boundaries. The interest of the method is to be simple since the damping model is already available in most of the general purpose finite element softwares.

## 2. A Simple Multi-Directional Absorbing Layer Method

### 2.1 Basic idea

Since the better ideas to deal with spurious reflections are to consider absorbing layers and multi-directional attenuating properties, our goal is to proceed as for the PML technique but with a simple description of the attenuation process. Physical models are generally very efficient to describe the attenuation process (frequency dependence, causality, etc) but they are nevertheless not very easy to implement or cost effective due to the use of memory variables for instance [35,36,37,38,39,40,41]. Since Caughey damping is already available in most of the finite element softwares, we propose to use this formulation to describe the attenuation of the waves in an absorbing layer of finite thickness. The method is thus called the *Caughey Absorbing Layer Method* or "CALM". This damping formulation is purely numerical but, as recalled hereafter, a rheological interpretation is possible in some cases (i.e. 2<sup>nd</sup> order Caughey formulation).

### 2.2 Rayleigh and Caughey damping

Rayleigh damping is a classical method to easily build the damping matrix [C] for a Finite Element model [3,4] under the following form:

$$[C] = a_0[M] + a_1[K] \quad (4)$$

where [M] and [K] are the mass and stiffness matrices of the whole model respectively. It is then called *Rayleigh damping matrix* and $a_0, a_1$ are the Rayleigh coefficients. [C] is the sum of two terms: one is proportional to the mass matrix, the other to the stiffness matrix.

A more general damping formulation was proposed by Caughey [42,43] and is expressed as follows:

$$[C] = [M]\sum_{j=0}^{m-1} a_j \left([M]^{-1}[K]\right)^j \quad (5)$$

As evidenced by Eq. (5), the Rayleigh formulation (Eq. (4)) corresponds to a 2$^{nd}$ order Caughey damping ($m$=2) involving a linear combination of the mass and stiffness matrices.

These ways to build the damping matrix are very convenient since it can be easily computed and is often available in general purpose FEM softwares. Furthermore, for modal approaches, the *Rayleigh (or Caughey) damping matrix* is diagonal in the real modes base [4,44]. Damping is therefore called proportional or classical. In case of non proportional damping, the complex modes have to be computed (in order to uncouple the modal equations).

Rayleigh (or Caughey) damping formulation may also be used to analyze the propagation of damped elastic wave [4,45]. For such problems, it may be useful to have a rheological interpretation of these purely numerical formulations. In the field of mechanical wave propagation, the equivalence between the Rayleigh formulation and a Generalized Maxwell model was already proposed [45] and is briefly recalled hereafter.

## 2.3   *Rheological interpretation of Rayleigh damping*

Considering Rayleigh damping [4,44] (i.e. 2$^{nd}$ order Caughey damping), the loss factor $\eta$ can be written as follows:

$$\eta = 2\xi = \frac{a_0}{\omega} + a_1 \omega \tag{6}$$

where $\omega$ is the circular frequency and $\xi$ is the damping ratio.

Considering the relationship between internal friction and frequency for Rayleigh damping, it is possible to build a rheological model involving the same attenuation-frequency dependence [45]. For a linear viscoelastic rheological model of complex modulus $M=M_R+iM_I$ [4,46], the expression of the quality factor $Q$ is given in the fields of geophysics and acoustics as follows:

$$Q = \frac{M_R}{M_I} \tag{7}$$

For weak to moderate Rayleigh damping, there is a simple relation between the inverse of the quality factor $Q^{-1}$ and the damping ratio $\xi$ [4,44]:

$$Q^{-1} \approx 2\xi \tag{8}$$

For Rayleigh damping, the loss factor is infinite for zero and infinite frequencies. It clearly gives the behaviour of the model through instantaneous and long term responses. As shown in [45], the rheological model perfectly meeting these requirements (attenuation-frequency dependence, instantaneous and long term effects) is a particular type of generalized Maxwell model.

Fig. 2 gives a schematic of the proposed rheological interpretation: it connects, in parallel, a classical Maxwell cell to a single dashpot. The generalized Maxwell model may be characterized by its complex modulus [4,46] from which we easily derive the inverse of the quality factor $Q^{-1}$ which takes the same form as the loss factor of Rayleigh damping (expression (6)) : it is the sum of two terms, one proportional to frequency and one inversely proportional to frequency.

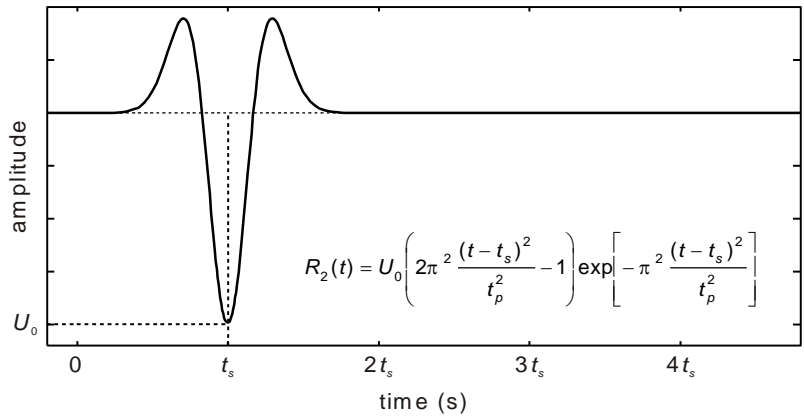

**Fig. 1.** 2$^{nd}$ order Ricker wavelet $R_2(t)$: maximum amplitude $U_0$, time shift $t_s$ and fundamental period $t_p$ [4,49].

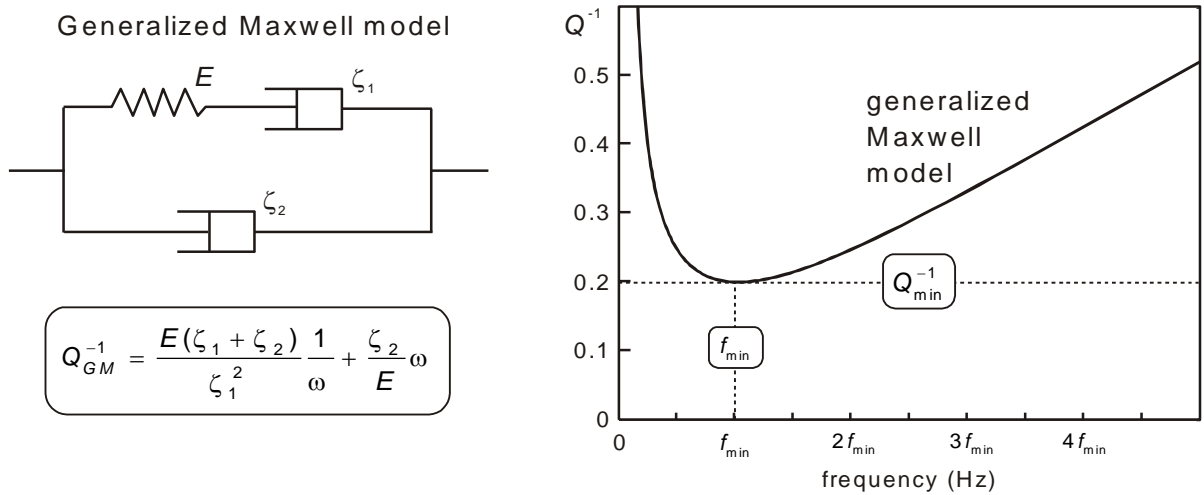

**Fig. 2.** Generalized Maxwell model and corresponding loss factor curve for longitudinal waves in a bar [45].

Considering Eq. (6) and Fig. (2), the Rayleigh coefficients can be easily related to the rheological parameters of the Generalized Maxwell model:

$$\begin{cases} a_0 = \dfrac{E(\zeta_1 + \zeta_2)}{\zeta_1^2} \\ a_1 = \dfrac{\zeta_2}{E} \end{cases} \quad (9)$$

This equivalence between Rayleigh damping and the Generalized Maxwell model is evidenced for wave propagation problems in [45]. If one needs to characterize a real material, the Rayleigh coefficients may thus be identified from experimental results. In this paper, a target theoretical damping will be chosen for a fictitious absorbing material (no experiments are needed).

### *2.4 Caughey Absorbing Layer Method*

In the framework of the Finite Element Method, an elastic medium is considered and an absorbing layer system will be designed at its boundaries. The absorbing layer is thus modelled with appropriate damping properties (i.e. Rayleigh/Caughey damping coefficients) in order to attenuate the spurious reflections at the mesh boundaries. This *Caughey Absorbing Layer Method* (*"CALM"*) may thus reduce the amplitude of the elastic wave coming from the elastic medium and reflecting at the artificial boundaries of the medium. In the following, the proposed technique is described for different damping variations in the absorbing layer thickness. The spatial variations of damping are controlled by variable damping coefficients in the Finite Elements. Such techniques were already used to model wave propagation in media with stress state dependent damping [47,48]. Considering Rayleigh damping, the element damping matrix for finite element "e" is thus written:

$$[C]^{(e)} = a_0^{(e)}[M]^{(e)} + a_1^{(e)}[K]^{(e)} \tag{10}$$

where $[M]^{(e)}$ and $[K]^{(e)}$ are the mass and stiffness matrices for finite element "e" respectively. The Rayleigh damping coefficients $a_0^{(e)}$ and $a_1^{(e)}$ may be different in each finite element or chosen piecewise constant in the absorbing layer. In the following, the efficiency of the *Caughey Absorbing Layer Method* is assessed for 1D and 2D elastic wave propagation.

## 3. Efficiency of the 1D Caughey Absorbing Layer

### *3.1 Definition of the propagating wave*

For the 1D case, a 2$^{nd}$ order Ricker wavelet is considered [4,49]. As depicted in Fig. 1, this type of wavelet is derived from a Gaussian and is rather well localized in both time and frequency domain [4]. It is thus ideal to perform a detailed analysis on wave propagation in a narrow frequency band investigating the reflection of short duration waves (to easily distinguish incident and reflected waves). We shall consider longitudinal elastic waves first and the predominant frequency of the Ricker wavelet $f_R = 1/t_p$ will be chosen in order to have an integer number of wavelengths along the medium. For the finite element computations, a Newmark non dissipative time integration scheme is chosen in order to avoid algorithmic damping [4,12] due to the time integration process.

### *3.2 Rayleigh damping in the absorbing layer*

In the following, the absorbing layer involves Rayleigh damping (2$^{nd}$ order Caughey damping) which is frequency dependent (Fig.2) and is rheologically equivalent to a Generalized Maxwell model for wave propagation problems [45]. To define a reference attenuation value (inverse of the quality factor $Q^{-1}$) in the absorbing layer, the minimum attenuation value will be chosen at the predominant frequency of the Ricker wavelet $f_R$. Since the Generalized Maxwell model has a band-pass behavior (Fig.2, right), it is thus probably the worst case in terms of efficiency of the absorbing layer. From the expression of the damping ratio, Eq. (6), it is possible to determine the frequency of minimum damping $\omega_{min}$ from the Rayleigh coefficients as follows:

$$\omega_{min} = \sqrt{\frac{a_0}{a_1}} \tag{11}$$

Choosing the minimum damping $\xi_{min}$ (or attenuation $Q_{min}^{-1}$) at the predominant frequency of the Ricker wavelet $f_R$, it is then possible to derive the following relation:

$$\omega_R = 2\pi f_R = \sqrt{\frac{a_0}{a_1}} \tag{12}$$

and thus, using the definition of Rayleigh damping, derive the following system:

$$\begin{cases} Q_{min}^{-1} = 2\xi_{min} = \dfrac{a_0}{\omega_R} + a_1 \omega_R \\ \omega_R = 2\pi f_R = \sqrt{\dfrac{a_0}{a_1}} \end{cases} \tag{13}$$

From the choice of the predominant frequency of the Ricker wavelet $f_R$ and the minimum attenuation $Q_{min}^{-1}$, the Rayleigh damping coefficients in the absorbing layer may then be estimated. In the following, we shall choose several typical values for $Q_{min}^{-1}$ ranging from 0.5 (i.e. $\xi_{min}$=0.25 or 25%) to 2.0 (i.e. $\xi_{min}$=1.0 or 100%).

### 3.3 Homogeneously absorbing case

#### 3.3.1 Description of the homogeneous absorbing layer system

As depicted in Fig. 3, the first numerical case corresponds to a 1D elastic medium (left) and a homogeneously absorbing layer (right). The length of the elastic layer (left) is $4\lambda$ and that of the homogeneously damped layer (right) is $\lambda$ ($\lambda$: wavelength of the longitudinal wave). Linear quadrilateral finite elements are considered and a $\lambda/20$ size is chosen in order to have low numerical wave dispersion [11,12]. In the elastic medium, the element damping matrices $[C]^{(e)}$ are zero whereas homogeneous Rayleigh damping is considered in the absorbing layer by choosing identical Rayleigh coefficients $a_0^{(e)}, a_1^{(e)}$ for each element damping matrix in this area (the elastic properties being identical in both domains). In each case, the attenuation value $Q^{-1}$ is chosen as the minimum attenuation value at the predominant frequency of the propagating wave (Ricker wavelet). As shown in Fig. 2, because of the damping-frequency dependence given by the Rayleigh formulation, all other frequency components are thus more strongly attenuated in the layer. Three attenuation values were chosen for the homogeneously absorbing case: $Q_{min}^{-1}$ =0.5, 1.0 and 2.0 (leading to damping values $\xi_{min}$ =25, 50 and 100% respectively).

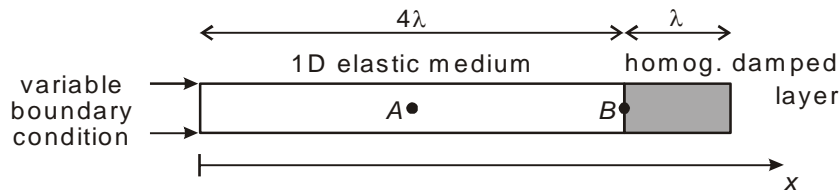

**Fig. 3.** Schematic of the first numerical test: undamped elastic layer (left) and homogenously damped layer (right).

The equation of motion of the whole finite element model is thus:

$$[M]\{\ddot{u}\} + [C]\{\dot{u}\} + [K]\{u\} = \{0\} \tag{14}$$

with the following variable boundary condition at $x$=0:

$$u_\ell(x=0, t) = R_2(t, t_s, t_p) \tag{15}$$

[M], [C], [K] being the mass, damping and stiffness matrices (resp.), $u_\ell$ the $\ell^{th}$ component of displacement, $R_2(t,t_s,t_p)$ the Ricker wavelet and, as shown in Fig.1, $t_s, t_p$ its parameters (time shift and fundamental period resp.).

#### 3.3.2 Efficiency of the homogeneously damped layer

For the three different attenuation values in the absorbing layer, the results are displayed in Fig. 4 for point *A* (left) at the centre of the elastic medium and point *B* (right) at the interface between the elastic medium and the absorbing layer (see points location in Fig. 3).

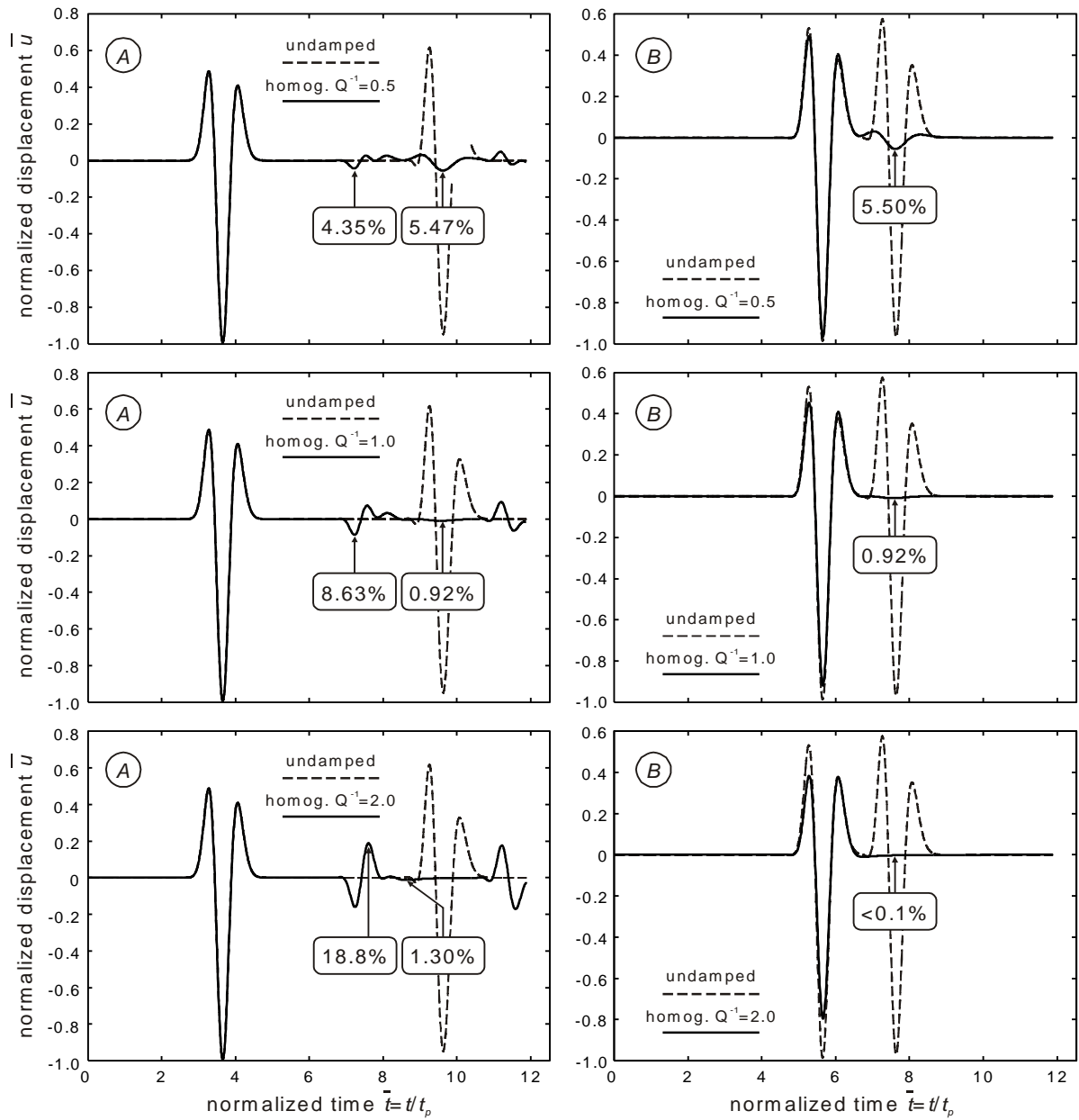

**Fig. 4.** Comparison between the homogeneously damped layer (solid) and the undamped case (dashed) at points *A* (left) and *B* (right) for different attenuations: $Q_{min}^{-1}$ =0.5 (top), $Q_{min}^{-1}$ =1.0 (middle) and $Q_{min}^{-1}$ =2.0 (bottom).

These curves are plotted in terms of normalized displacement $\bar{u} = u/U_0$ vs normalized time $\bar{t} = t/t_p$ ($U_0$ and $t_p$ being the Ricker wavelet amplitude and fundamental period resp.) and lead to the following conclusions:

- For $Q_{min}^{-1}$ =0.5 (top, solid), when compared to the undamped case ($Q_{min}^{-1}$ =0.0, dashed), the amplitude of the reflected wave at point *A* is much smaller (5.47% of $U_0$) but the incident wave is also reflected at the interface between the elastic medium and the absorbing layer (4.35%). It is due to the velocity contrast between the elastic medium and the viscoelastic layer (in terms of complex wavenumber). For $Q_{min}^{-1}$ =0.5 at point *B* (top right, solid), the amplitude at the interface between the elastic medium and the absorbing layer is also small (5.50%). The efficiency of the homogeneously absorbing layer thus appears acceptable.

- For $Q_{min}^{-1}$ =1.0 (middle, solid), the amplitude of the reflected wave at the end of the absorbing layer is very small (0.92% of $U_0$) but the reflected wave at the interface with the elastic medium is larger than for $Q_{min}^{-1}$ =0.5 (8.63% instead of 4.35%). This is due to the fact that the complex velocity contrast with the elastic medium is larger for $Q_{min}^{-1}$ =1.0.
- Finally, for $Q_{min}^{-1}$ =2.0 (bottom, solid), the contrast being larger again, the results are not very good for the wave reflected at the interface (18.8%). However the reflected wave at the end of the absorbing layer is again very small (1.30%) and the amplitude at the interface (point *B*) is nearly zero.

The authors obtained similar results for transverse waves (*SV* waves). From these three different homogeneously damped cases, the *Caughey Absorbing Layer Method* (*"CALM"*) can already be considered as an efficient absorbing layer method but its efficiency may probably be improved and its artefacts reduced.

### *3.4 Heterogeneously absorbing case*

In the homogeneous case, the velocity contrast between the elastic layer and the absorbing layer may have a detrimental effect. The idea is now to have an increasing damping value in the absorbing layer system along the direction of the incident wave and a lower damping contrast at the interface with the elastic medium.

#### 3.4.1 Description of the heterogeneous absorbing layer system

As depicted in Fig. 5, the second numerical case corresponds to a heterogeneously absorbing layer. The absorbing layer is divided into two λ/2 thick (top) or five λ/5 thick sub layers (bottom) involving piecewise constant Rayleigh damping coefficients in each sub-layer but increasing from one layer to the other. Two different sets of minimum attenuation values in each absorbing sub-layer are chosen:

- 1$^{st}$ set: $Q_{min}^{-1}$ =1.0 in the rightmost sub-layer and piecewise constant in each other sub-layers ($Q_{min}^{-1}$ =0.5 in the first layer of the two-layers case and $Q_{min}^{-1}$ =0.2, 0.4, 0.6 and 0.8 for the leftmost layers in the five-layer case).
- 2$^{nd}$ set: $Q_{min}^{-1}$ =2.0 in the rightmost sub-layer and piecewise constant in each other sub-layers ($Q_{min}^{-1}$ =1.0 in the first layer of the two-layers case and $Q_{min}^{-1}$ =0.4, 0.8, 1.2 and 1.6 for the leftmost layers in the five-layer case).

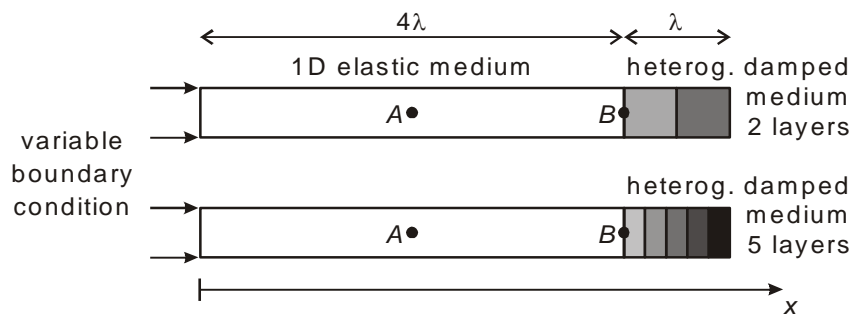

**Fig. 5.** Schematic of the second numerical test: undamped elastic layer (left) and heterogeneously damped layer (right) involving 2 layers (top) or 5 layers (bottom)

### 3.4.2 Efficiency of the heterogeneously damped layer

For the heterogeneously damped case, the results are displayed in Fig. 6 for point *A* (i.e. centre of the elastic medium) and compared to the undamped case and the homogeneously damped case (1st set of damping values: top, 2nd set: bottom). A closer view on the reflected waves is also proposed (right). Since the incident elastic wave is reflected at the interface and at the absorbing layer boundary, two different relative amplitudes are obtained for both reflected waves.

For the 1st set of damping values (i.e. $Q_{min}^{-1}$ =1.0 in the rightmost sub-layer):

- The first value corresponds to the reflection at the interface and ranges from 8.63% for the homogeneous case down to 1.66% for the five-layers case (4.35% for 2 layers).
- The second value is related to the reflection at the absorbing layer boundary and ranges from 0.92% for the homogeneous case up to 4.41% for the five-layers case (2.10% for 2 layers).

For the 2nd set of damping values (i.e. $Q_{min}^{-1}$ =2.0 in the rightmost sub-layer):

- The first value corresponds to the reflection at the interface and ranges from 18.8% for the homogeneous case down to 2.78% for the five-layers case (8.63% for 2 layers).
- The second value is related to the reflection at the absorbing layer boundary and ranges from 1.30% for the homogeneous case up to 2.86% for the five-layers case (2.83% for 2 layers).

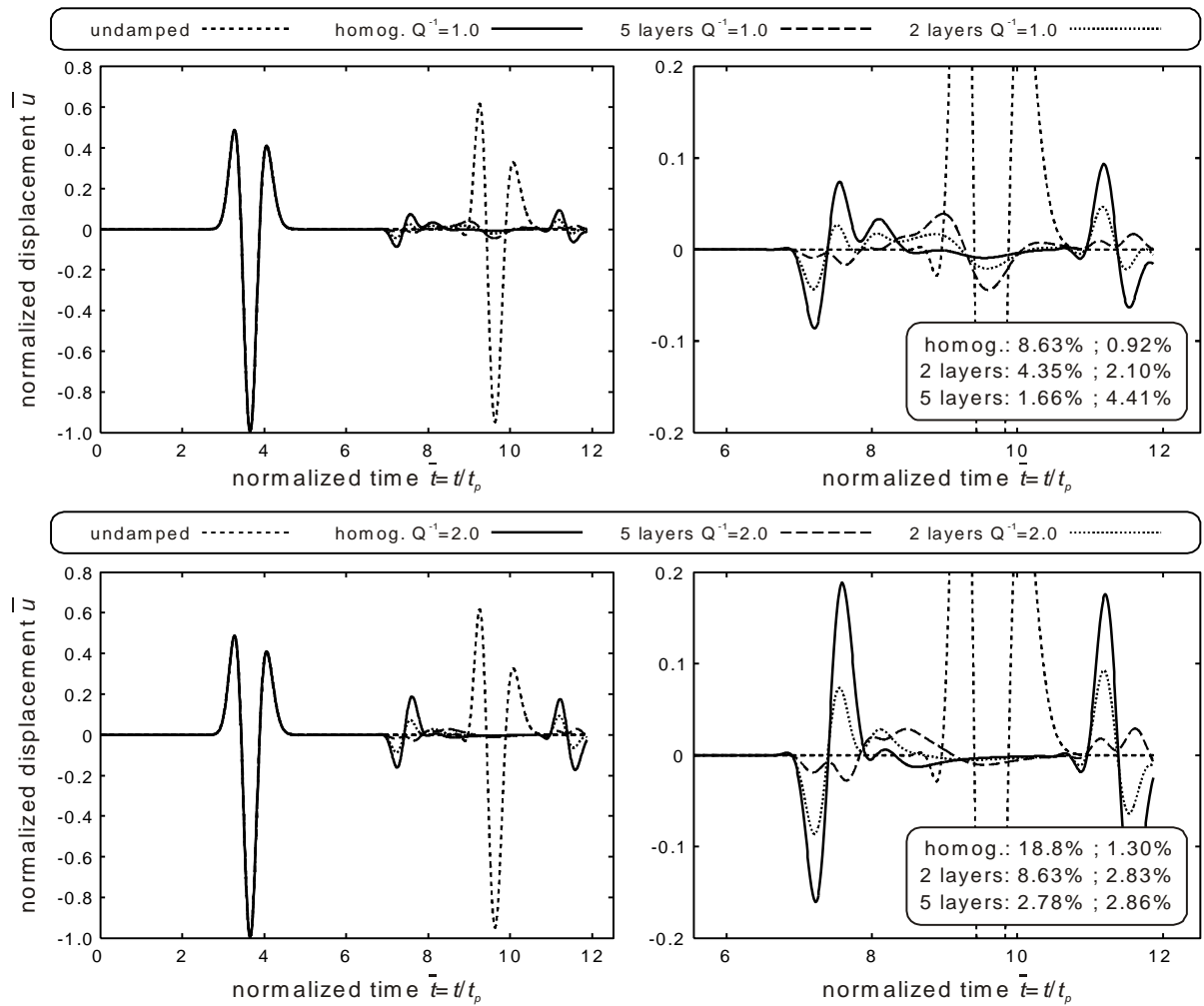

**Fig. 6.** Comparison between the heterogeneously damped layer (dashed or dotted) and the homogeneous case (solid) at point *A*: 1st set (top) and 2nd set of attenuation values (bottom).

From these results, it appears that it is necessary to balance the advantage of a strong damping in a thick layer (strong amplitude reduction in the layer) and slow variations of the damping properties (low amplitude for the reflections at the interfaces). Since in the PML approach, the parameters governing the amplitude decrease (or the coordinate stretching) are "perfectly matched" at the interface [26,27,28,29,30], the case of a continuous damping variation in the absorbing layer will thus be studied hereafter.

### *3.5 Continuously varying damping*

#### 3.5.1 Description of the continuous absorbing layer system

As depicted in Fig. 7, the third numerical case corresponds to a continuously varying damping in the absorbing layer. The absorbing layer involves variable Rayleigh damping coefficients increasing linearly with the horizontal distance. The idea is to have a continuously increasing damping value in the absorbing layer system. Two cases are considered: attenuation $Q_{min}^{-1}$ increasing linearly from 0 to 1.0 (1$^{st}$ case) and from 0 to 2.0 (2$^{nd}$ case).

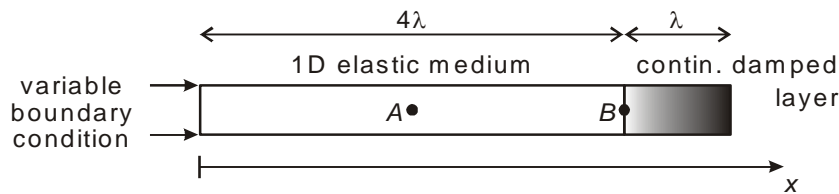

**Fig. 7.** Schematic of the third numerical test: undamped elastic layer (left) and continuously absorbing layer system (right).

#### 3.5.2 Efficiency of the continuously absorbing layer system

For the continuously (i.e. linearly) damped case, the results at point *A* (centre of the elastic medium) are displayed in Fig. 8 for the two cases (top/bottom) and compared to the undamped case ($Q_{min}^{-1}$ =0.0). Closer views are also displayed in this Figure (right). For both cases, the efficiency of the continuous absorbing layer system is slightly better than that of the 5-layers system for the reflection at the interface: 1.11% instead of 1.66% for $Q_{min}^{-1}$ =1.0 (top), 2.17% instead of 2.78% for $Q_{min}^{-1}$ =2.0 (bottom). For the reflection at the medium boundary, the efficiency of the continuous system is a bit less for $Q_{min}^{-1}$ =1.0 (5.25% vs 4.41%) whereas it is nearly the same for $Q_{min}^{-1}$ =2.0 (2.80% vs 2.86%). When compared to the homogeneous case, the overall efficiency of the 5-layers and continuous systems are satisfactory.

### *3.6 Influence of the size of the absorbing layer*

#### 3.6.1 Description of the alternative absorbing layer system

Since the amplitude decay is influenced by the travelling distance, it is necessary to assess the influence of the absorbing layer thickness, another configuration involving a half wavelength thick layer is thus considered (Fig. 9 bottom). The thickness of this absorbing layer system being smaller than in the previous case (Fig. 9 top), the efficiency of this configuration should be less due to a shorter travelling distance in the layer (see Eq. (3)) whereas the number of degrees of freedom in the Finite Element model will be less.

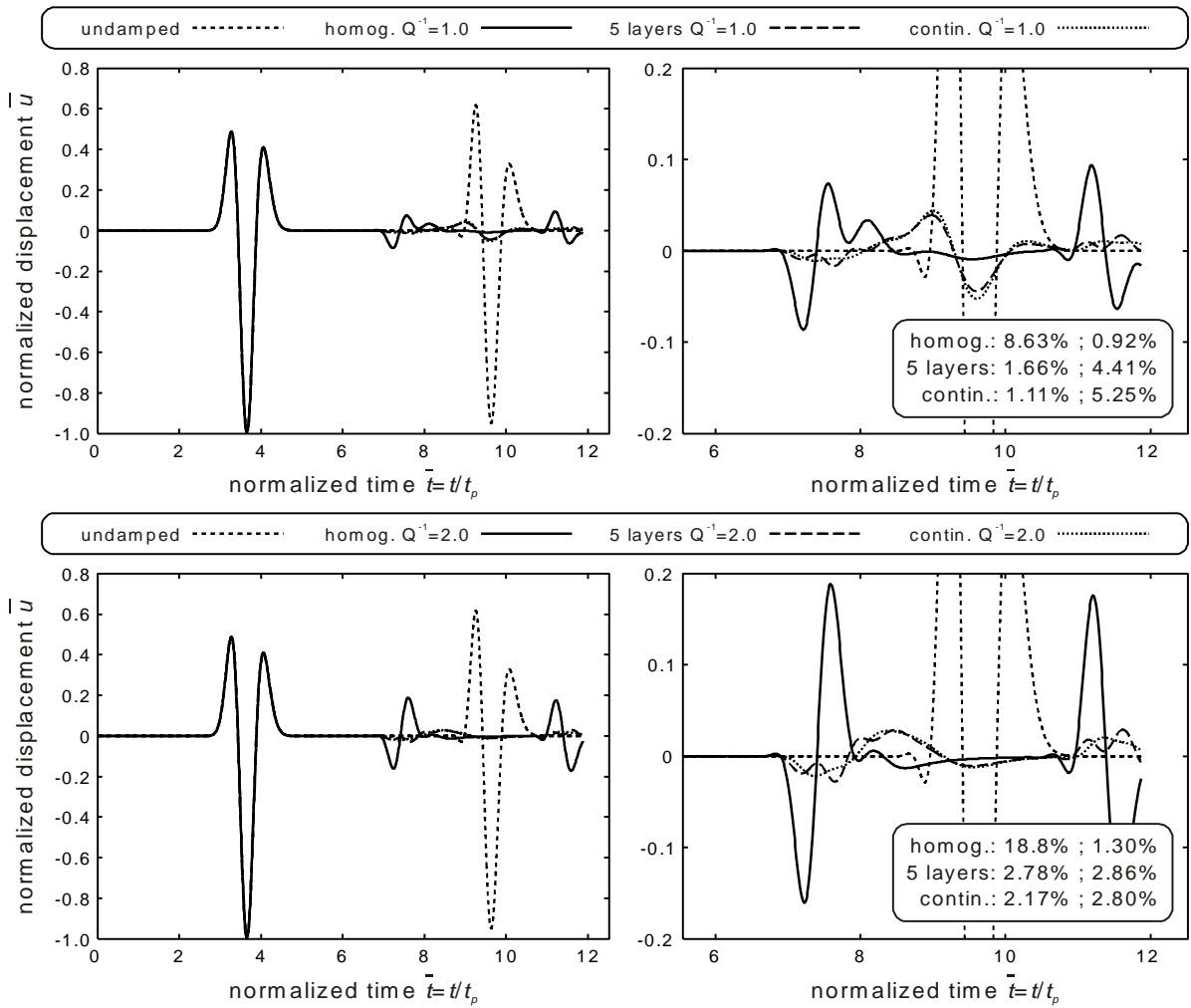

**Fig. 8.** Comparison between the continuously damped layer and the homogeneous, heterogenous and undamped cases at point *A*: 1st case (top) and 2nd case (bottom). Closer views at right.

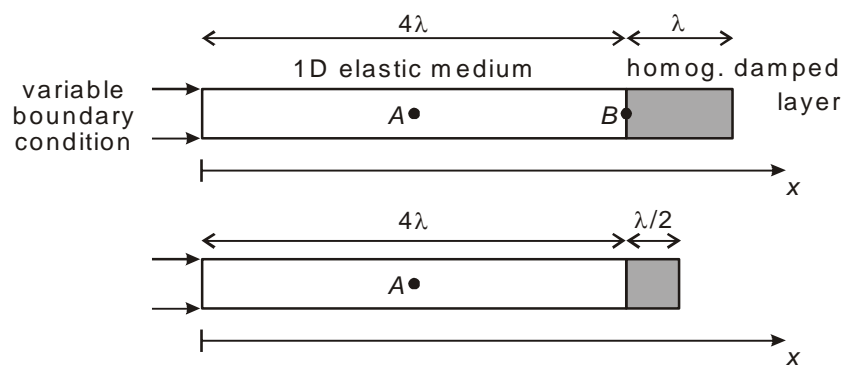

**Fig. 9.** Various absorbing layer thicknesses to compare their efficiency: one wavelength thick layer (top) and half wavelength thick layer (bottom).

### 3.6.2 Influence of the thickness of the absorbing layer

The influence of the absorbing layer thickness is shown in Fig. 10 for the homogeneous case (top) and the continuous case (bottom) and for two different attenuations $Q_{min}^{-1}$ =1.0 (left) and $Q_{min}^{-1}$ =2.0 (right).

For the reflection at the interface, the results are:
- *Homogeneous case (top)*: identical efficiency (8.63% for $Q_{min}^{-1}$ =1.0 and 18.8% $Q_{min}^{-1}$ =2.0) for both thicknesses since the velocity (or complex modulus) contrast is the same.
- *Continuous case (bottom)*: since the linear increase of damping is faster for the thinnest layer, the velocity contrast is a bit larger thus leading to a larger amplitude (2.19% instead of 1.11% for $Q_{min}^{-1}$ =1.0 and 4.21% instead of 2.80% for $Q_{min}^{-1}$ =2.0).

For the reflection at the model boundary, the efficiency of the half wavelength absorbing layer is significantly less than that of the one wavelength case since the distance along which the waves are attenuated is much less:
- *Homogeneous case (top)*: for $Q_{min}^{-1}$ =1.0 (left), the relative amplitude of the reflected wave is 6.80% for the $\lambda/2$ thick absorbing layer instead of 0.92% for the $\lambda$ thick layer. For $Q_{min}^{-1}$ =2.0, the relative amplitude is 2.33% for the $\lambda/2$ thick instead of 1.30%.
- *Continuous case (bottom)*: for $Q_{min}^{-1}$ =1.0 (left), the relative amplitude of the reflected wave is 16.9% for the $\lambda/2$ thick absorbing layer instead of 5.25% for the $\lambda$ thick layer. For $Q_{min}^{-1}$ =2.0, the relative amplitude is 9.51% for the $\lambda/2$ thick instead of 2.80%.

The influence of the absorbing layer thickness (or length) is thus very large and $\lambda/2$ is obviously not a very efficient choice since the efficiency is much less than for a $\lambda$-thick absorbing layer and the relative reduction of the number of degrees of freedom in the finite element model would not be so large for a wide model.

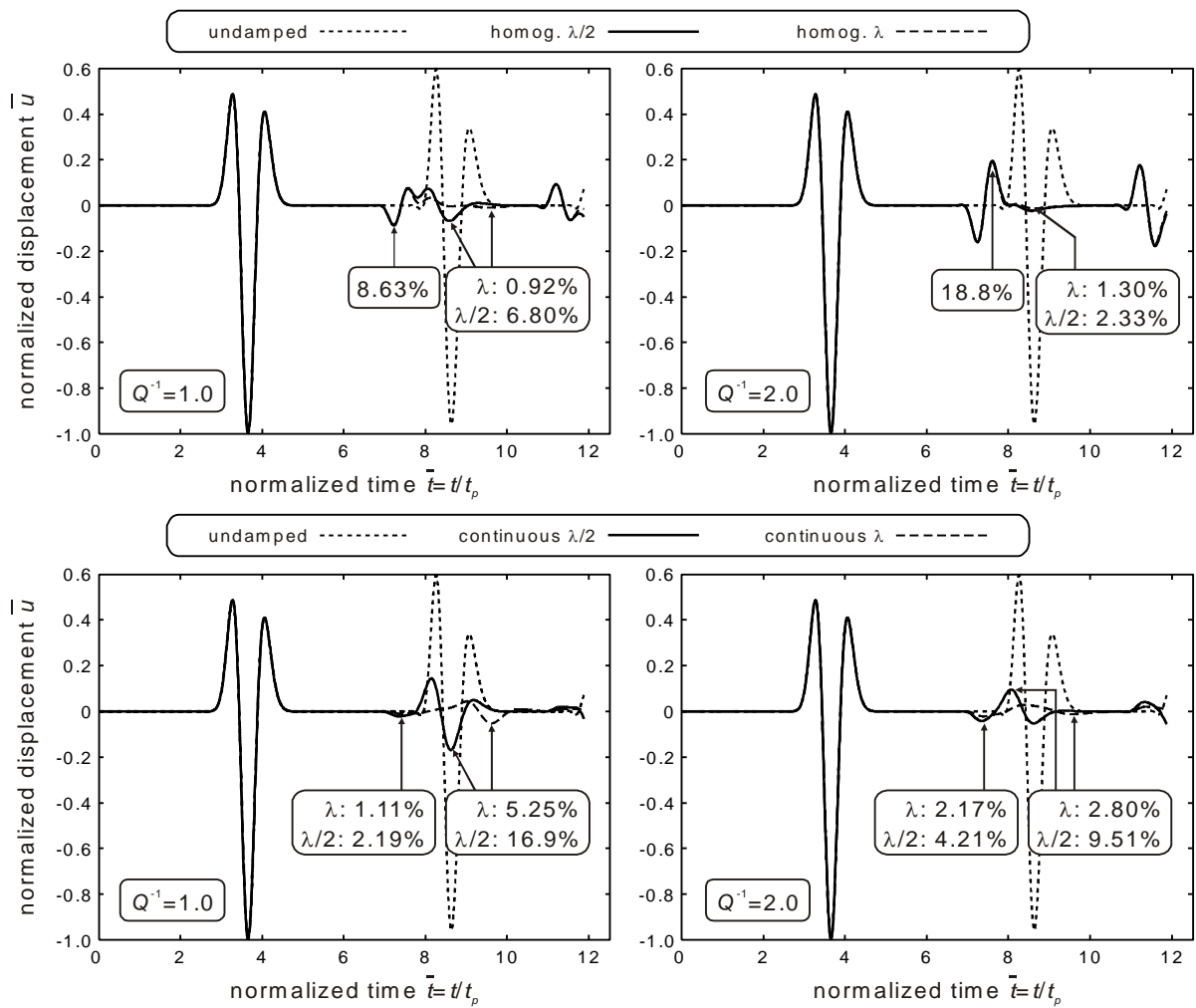

**Fig. 10.** Influence of the thickness of the absorbing layer system: homogeneous case (top), continuous case (bottom).

## 4. Efficiency of the 2D Caughey Absorbing Layer for pure P-waves

### *4.1 Simple 2D case from the PML technique: geometry and loading*

To analyze the efficiency of the *CALM* in 2D configurations, we shall first study a simple 2D case involving pure *P*-waves. We shall consider a simple 2D case proposed by Festa and Vilotte [22] for pure *P*-waves in the framework of the PML technique. As shown in Fig. 11, a 2D absorbing layer is designed all around a square elastic domain. The wave velocities are $V_P$=4 km/s and $V_S$=2.31 km/s and the mass density is $\rho$=2.5 g/cm$^3$ (see [22] for other details on the PML model). The size of the elastic domain (2L=1500m) thus corresponds to 3.75$\lambda_P$ and the width of the Caughey absorbing layer is chosen as $\lambda_P$/2. As Festa and Vilotte [22] chose an explosive source using a Ricker wavelet for the time variations ($t_p$=0.1s, i.e. $f_p$=10Hz), we considered a finite element model (Fig. 11 right) involving a pressure loading with similar time variations.

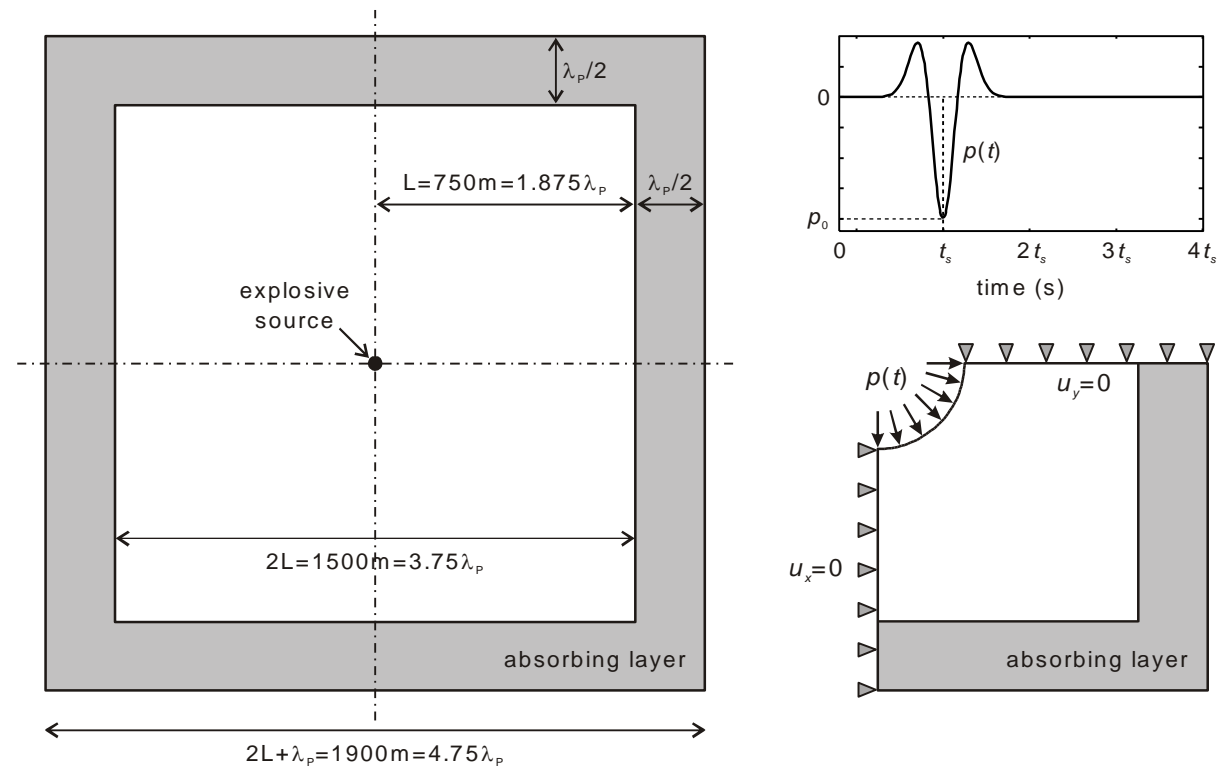

**Fig. 11.** 2D plane strain model for pure P-waves proposed by Festa and Vilotte (2005): adapted schematic (left), pressure loading (top right) and finite element model considered for the *CALM* (bottom right).

### *4.2 Comparison between the Caughey Absorbing Layer Method and PMLs*

The wavefield is computed in terms of normalized displacement $\bar{u} = \|u\|/|p_0|$ ($\|u\|$ being the norm of the displacement vector and $|p_0|$ the maximum loading pressure) for the undamped 2D case and the 5 layers 2D case ($Q_{min}^{-1}$ =0.4, 0.8, 1.2, 1.6 and 2.0 in each layer resp.). The numerical results are displayed in Fig. 12 for three different normalized times $\bar{t}_1 = t_1/t_p = 2.49$ (left) ; $\bar{t}_2 = t_2/t_p = 3.32$ (centre) ; $\bar{t}_3 = t_3/t_p = 4.56$ (right). For the undamped case (Fig. 12 top), the wave reflections at the model boundaries are obvious for times $\bar{t}_2$ and $\bar{t}_3$. For the Caughey absorbing layer (Fig. 12 bottom), the wavefield is partially absorbed in the first layers at time $\bar{t}_2$ and no reflections from the model boundaries appear at time $\bar{t}_3$. Similar results were obtained by Festa and Vilotte [22] for the PML technique. The thickness of the absorbing layer is nevertheless larger in *CALM*. The efficiency is also less since, for the same configuration, Festa and Vilotte obtained reflections coefficients ranging from 0.1% to 1% for a cubic decay function and from 0.6% to 2% for a quadratic decay function [22]. They also studied 2D cases to describe soft geological deposits such as a thin surface layer or a curved surface layer.

Since the previous results correspond to a simple incident wavefield, we shall now study a more complex case in terms of both polarization and incidences.

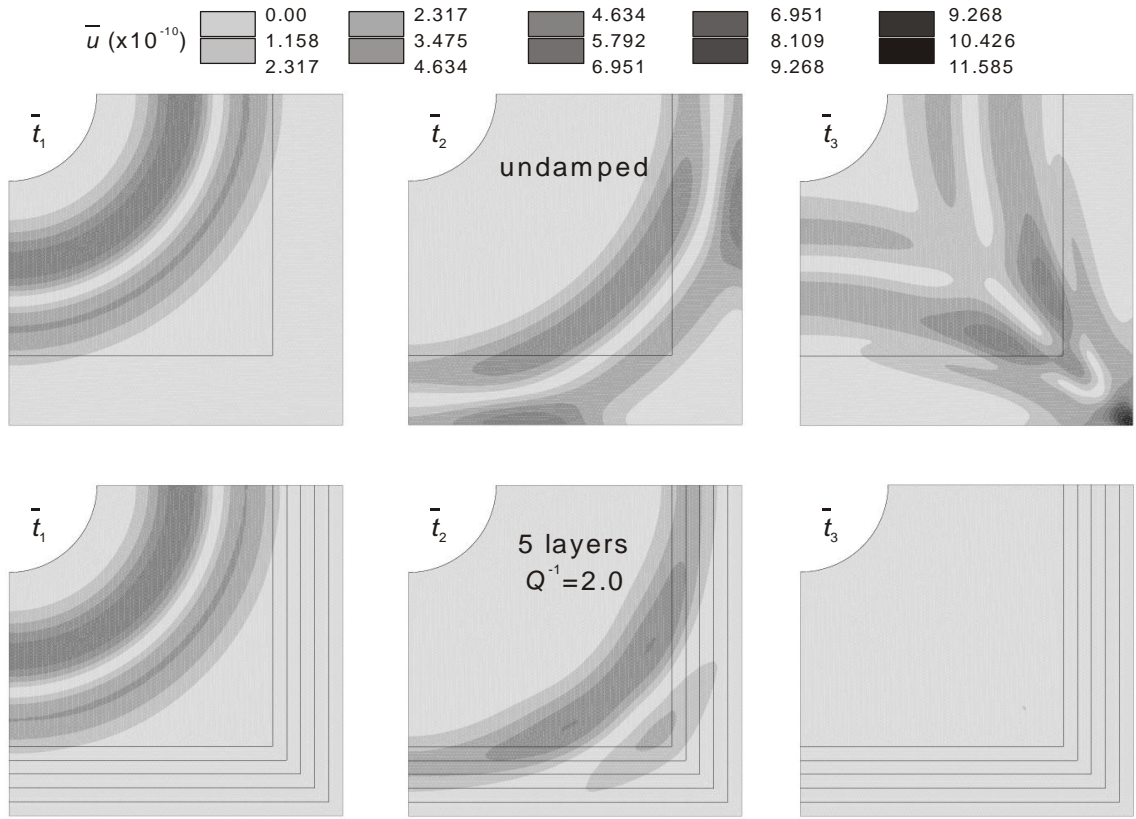

**Fig. 12.** Comparison between the undamped 2D case (top) and the 5 layers 2D case ($Q_{min}^{-1}$ =2.0, bottom) for pure P-waves at three different times: normalized displacement at $\bar{t}_1 = t_1 / t_p = 2.49$ (left) ; $\bar{t}_2 = t_2 / t_p = 3.32$ (centre) ; $\bar{t}_3 = t_3 / t_p = 4.56$ (right).

## 5. Efficiency of the 2D Caughey Absorbing Layer for various wave types

### *5.1 Definition of the propagating wave and geometry*

To assess the efficiency of the Caughey Absorbing Layer Method for more complex 2D wavefields, another 2D FEM model is considered (Fig. 13). A plane strain model is chosen in order to avoid strong geometrical damping. It involves a 4λx4λ square elastic medium and two λ thick absorbing layers (right and bottom). The model is symmetrical along the left boundary and the variable boundary condition (vertical displacement varying as a Ricker wavelet at $f_R$ =1/$t_p$) is applied at the free surface along a distance of λ/2 (Fig. 13). The wavefield in the model is thus composed of various wavetypes (longitudinal, transverse and surface waves) and the motion duration is larger than in the 1D case and the previous 2D case. The element size, λ/20, is identical to that of the 1D case.

It should be noticed that the total storage will be significantly increased but it is nevertheless independent from the model size and the relative storage amount will thus be smaller for a larger model.

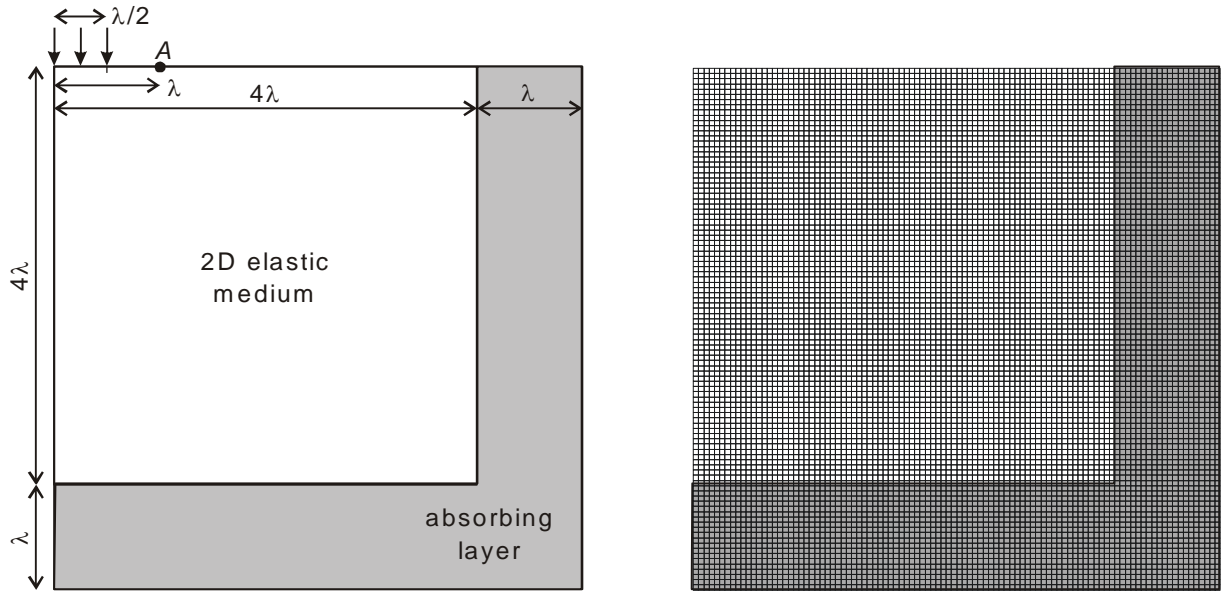

**Fig. 13.** Schematic of the 2D plane strain model (left) showing the boundary condition (top) and finite element mesh (right).

The efficiency of the homogeneous, heterogeneous and continuous damped cases will be assessed in the whole model (isovalues plots at some selected times) and at point *A* along the free surface (normalized displacement vs normalized time).

### *5.2 Efficiency of the homogeneous absorbing layers*

The results obtained in the homogeneous case are displayed in Fig. 14 in terms of normalized displacement (norm of the displacement vector divided by the maximum amplitude in the undamped case at the same time) at three different normalized times ($\bar{t}_1 = 4.15$, $\bar{t}_2 = 9.135$ and $\bar{t}_3 = 14.1$). For this displacement isovalue scale ($0.0 \leq \bar{u} \leq 0.6$), the reflected waves are obvious in the undamped case (top). For both homogeneously damped cases (middle and bottom), when compared to the amplitudes of the undamped case, the first reflections are not significant at time $\bar{t}_2$ (only the incident wavefield is present in this plot) and no reflection at all can be identified at time $\bar{t}_3$. The efficiency of the *Caughey Absorbing Layer Method* thus appears satisfactory in the 2D homogeneous case.

Since the displacement isovalues scale is rather large in Fig. 14, a narrower scale will now be considered in order to compare the various cases in details.

The numerical results at time $\bar{t}_3$ are displayed in Fig. 15 considering a displacement amplitude range $0.00 \leq \bar{u} \leq 0.02$ (some values above this range are displayed in white). In the homogeneous case (top), the relative amplitude of the reflected wave is small for $Q_{min}^{-1}$ =0.5 (2.48% of the maximum amplitude obtained in the undamped case (Fig. 15 top left)). For larger damping values ($Q_{min}^{-1}$ =1.0: top centre and $Q_{min}^{-1}$ =2.0: top right), the homogeneously absorbing layer system leads to larger amplitudes of the reflected waves (4.67% and 10.4% respectively) probably due to the velocity contrast at the interface (see results from the 1D case).

### *5.3 Efficiency of the heterogeneous absorbing layer*

For the heterogeneous case, five absorbing layers are considered at the medium boundaries (bottom and right). In order to assess the influence of the damping variations, the linear increase (already studied in the 1D case) is now compared to a quadratic and a square root law. The numerical results of the heterogeneous cases at time $\bar{t}_3$ are displayed in Fig. 15 (middle). The amplitude of the reflected wave is very small for the linear and square root laws (1.53% and 1.87% respectively) and a bit larger for the quadratic law (3.00%). The discrepancy is due to the slower increase of damping obtained with the latter.

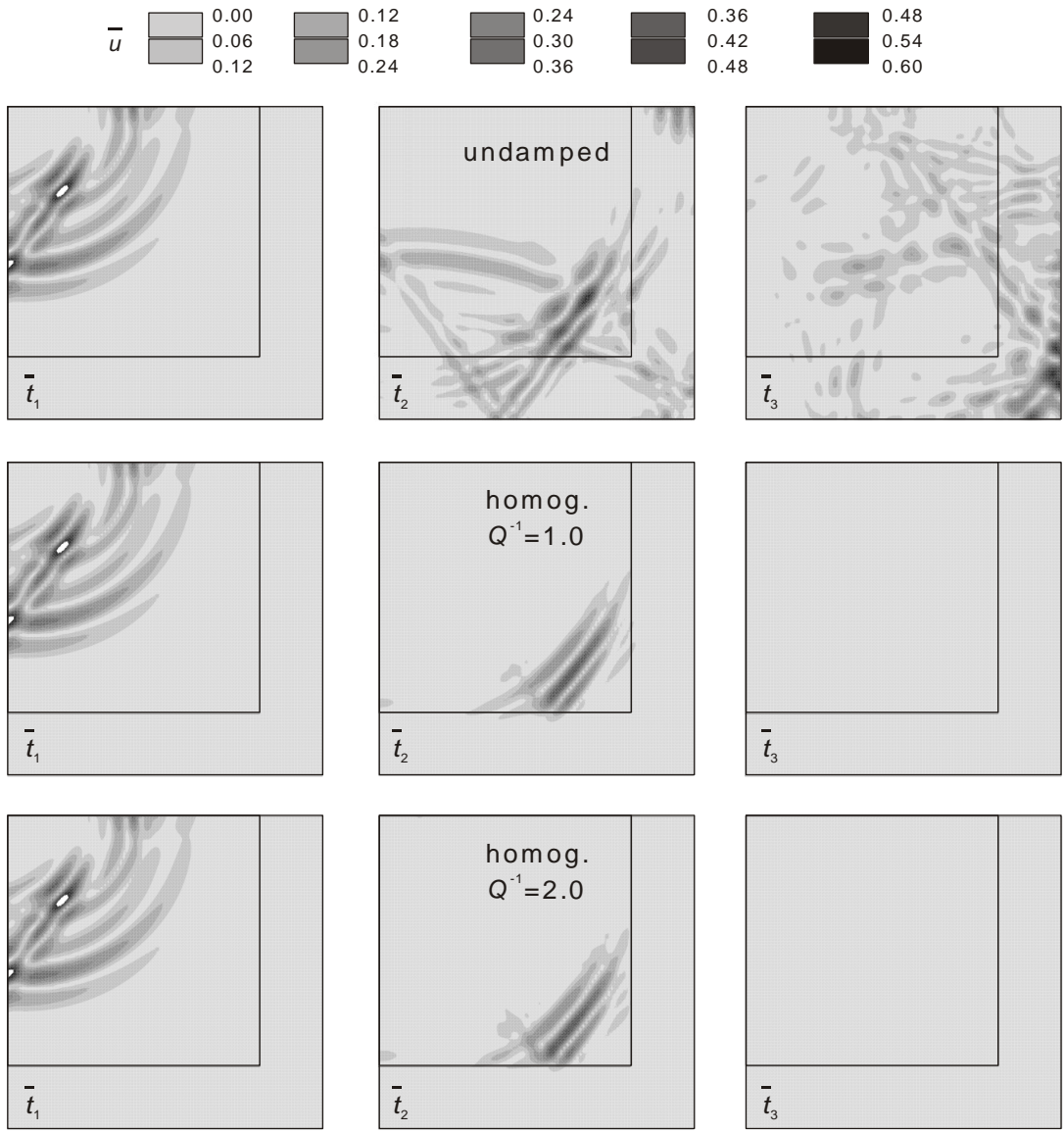

**Fig. 14.** Comparison between the undamped 2D case (top) and the homogeneous 2D case ($Q_{min}^{-1}$ =1.0, middle; $Q_{min}^{-1}$ =2.0, bottom) at three different times: normalized displacement at $\bar{t}_1 = t_1 / t_p = 4.15$ (left); $\bar{t}_2 = t_2 / t_p = 9.13$ (centre); $\bar{t}_3 = t_3 / t_p = 14.1$ (right).

## *5.4 Efficiency of the continuous absorbing layers*

For the continuous case, the linear increase of damping is again compared to a quadratic and a square root law. The numerical results of the continuous cases at time $\bar{t}_3$ are displayed in Fig. 15 (bottom). As for the heterogeneous case, the amplitude of the reflected wave is very small for the linear and square root laws (1.65% and 1.49% respectively) and a bit larger for the quadratic law (4.02%). The square root law leads to a faster damping increase near the interface and works better in the continuous case due to the regular increase of damping in the absorbing layer system (when compared to the piecewise constant damping in the heterogeneous case).

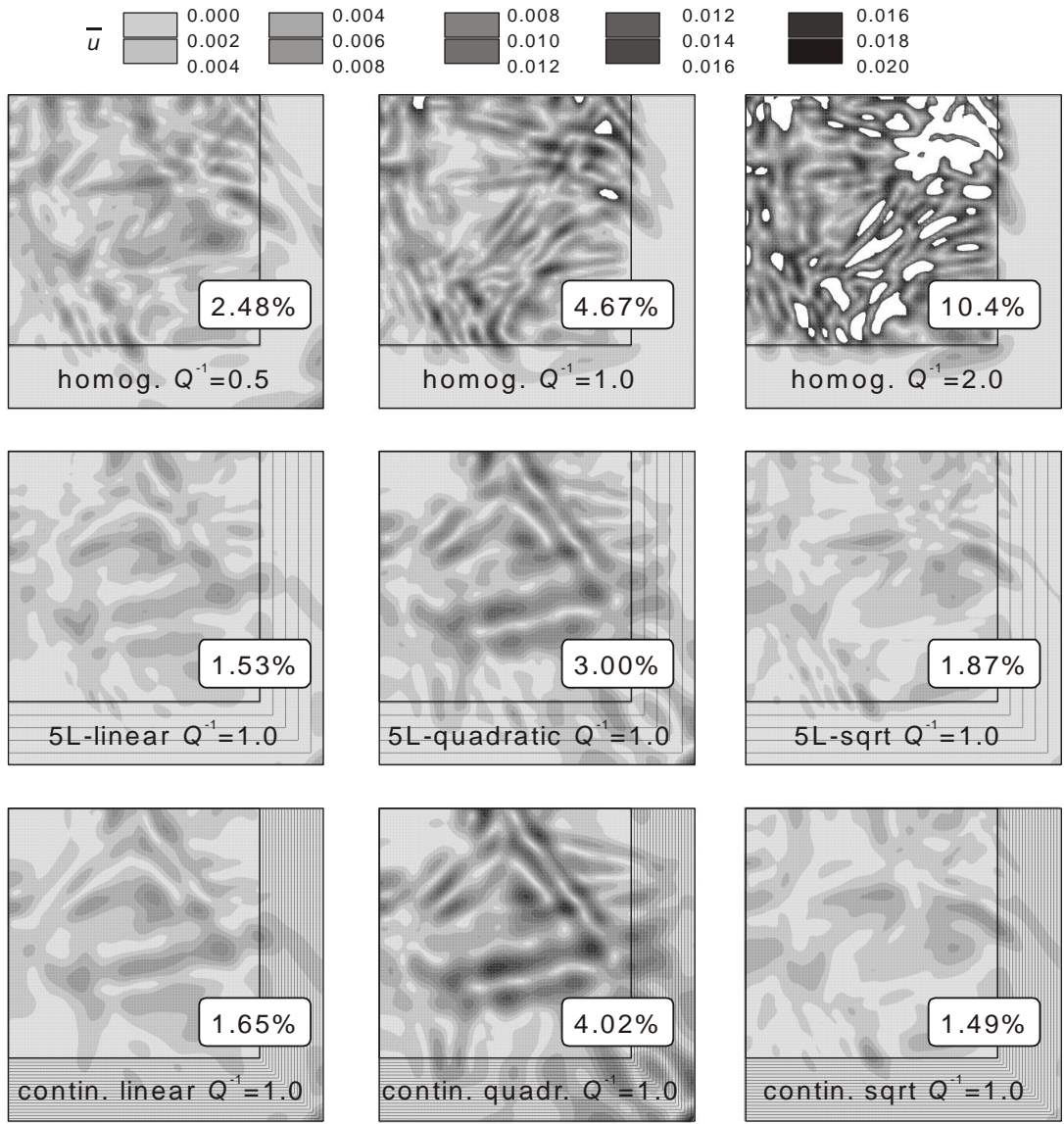

**Fig. 15.** Comparison between homogeneous (top), heterogeneous (5 layers, middle) and continuous (bottom) 2D cases at normalized time $\bar{t}_3 = t_3 / t_P = 14.1$. Maximum relative amplitude of the reflected waves in %.

### *5.5 Influence of the damping variations in the continuous layers*

To assess the influence of the damping variations in the continuous case for various maximum attenuation $Q^{-1}_{min}$, the maximum relative amplitude of the reflected waves are compared in Fig. 16. The homogeneous case (*a*) for $Q^{-1}_{min}$ =2.0 leads to a large amplitude (10.4%). For $Q^{-1}_{min}$ =1.0 (*b,c*), the values were already discussed in the two previous paragraphs (best solutions with the linear and the square root laws). From additional simulations with $Q^{-1}_{min}$ =2.0 (*d,e*), the square root law appears as the worst solution (3.39% and 2.66% resp.) since the quadratic case gives much better results (2.30% and 2.20% resp.) than for $Q^{-1}_{min}$ =1.0 (3.00% and 4.02% resp.). The continuous case (*e*) improves significantly the results for the different laws (when compared to the heterogeneous case (*d*)) due to the smoother description of the damping variations in the absorbing layer.

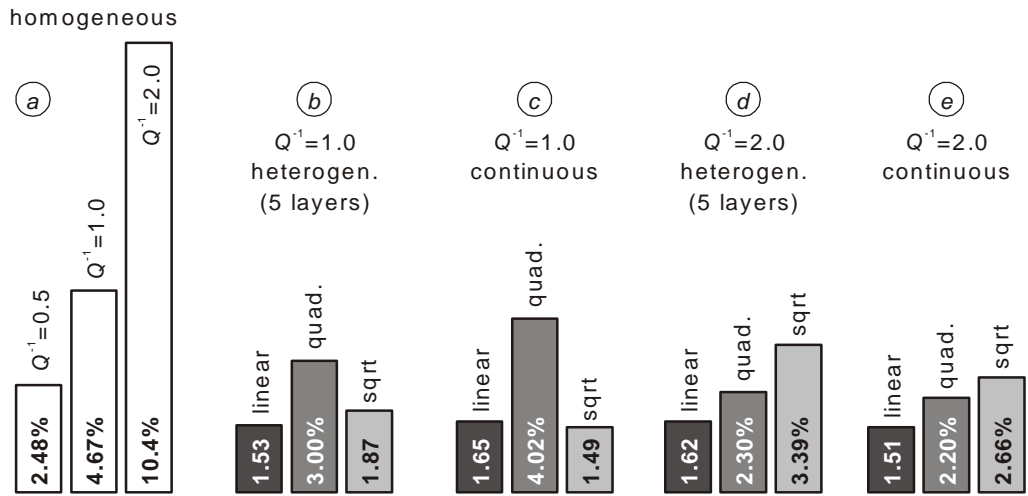

**Fig. 16.** Comparison between homogeneous 2D cases (*a*), heterogeneous (*b*,*d*) and continuous (*c*,*e*) 2D cases with various laws at normalized time $\bar{t}_3 = t_3 / t_P = 14.1$ in terms of maximum relative amplitude of the reflected waves.

### *5.6 Efficiency of Caughey Absorbing Layer for surface waves*

To assess the efficiency of the Caughey Absorbing Layer Method for surface waves, the numerical results at point *A* are now displayed in terms of normalized displacement vs normalized time (Fig. 17). All 2D cases are considered: homogeneous case (top); heterogeneous and continuous case for $Q_{min}^{-1}$ =1.0 (middle); heterogeneous and continuous case for $Q_{min}^{-1}$ =2.0 (bottom). As in the 1D case, the reflections at the interface and at the medium boundaries are quantified in terms of maximum relative amplitude.

In the homogeneous case (Fig. 17, top), the reflected surface waves have large amplitudes when compared to the undamped case (5.60% to 22.9%). In the heterogeneous and continuous case for $Q_{min}^{-1}$ =1.0 (Fig. 17, middle), the reflection at the interface leads to low amplitudes (2.21% and 1.21% resp.) whereas the reflections at the medium boundaries are a bit larger (3.90% and 5.01% resp.). As shown by the time histories, the variations with time are strong and it is difficult to quantify an overall maximum amplitude in both time (Fig. 17) and space (Fig. 15).

In the heterogeneous and continuous case for $Q_{min}^{-1}$ =2.0 (Fig. 17, bottom), the reflection at the interface leads to low amplitudes (4.11% and 2.54% resp.) and the reflections at the medium boundaries are also small (3.20% and 3.64% resp.). For such a maximum damping value, the continuous case is more efficient than the heterogeneous case. Since this 2D case involves various incidences, polarizations and wave types, the overall efficiency of the proposed method may thus be considered as satisfactory.

### 6. Conclusion

The main conclusion of this work is that the *Caughey Absorbing Layer Method* (at least the 2$^{nd}$ order *CALM* discussed herein) is reliable to reduce the spurious elastic wave reflections in Finite Element computations. Furthermore, when compared to *PML* techniques or viscoelastic mechanical models, it is very easy to implement (damping matrix directly computed from the stiffness and mass matrices) or even already available in most of the general purpose Finite Element softwares.

From the simple 2D simulations (pure P-waves), our results are at the same level as the 3$^{rd}$ order PML technique proposed by Festa and Vilotte [22] but our method needs a larger additional storage (thicker absorbing layers). For more complex 2D wavefields, the efficiency of the *Caughey Absorbing Layer Method* is also shown to be satisfactory. In such complex cases, the best efficiency of the *CALM* is reached for a damping variation up to $Q^{-1}_{min}$ =1.0 defined by a linear function in the heterogeneous case (5 layers with piecewise constant damping) and linear as well as square root function in the continuous case (1.53%, 1.65% and 1.49% resp.).

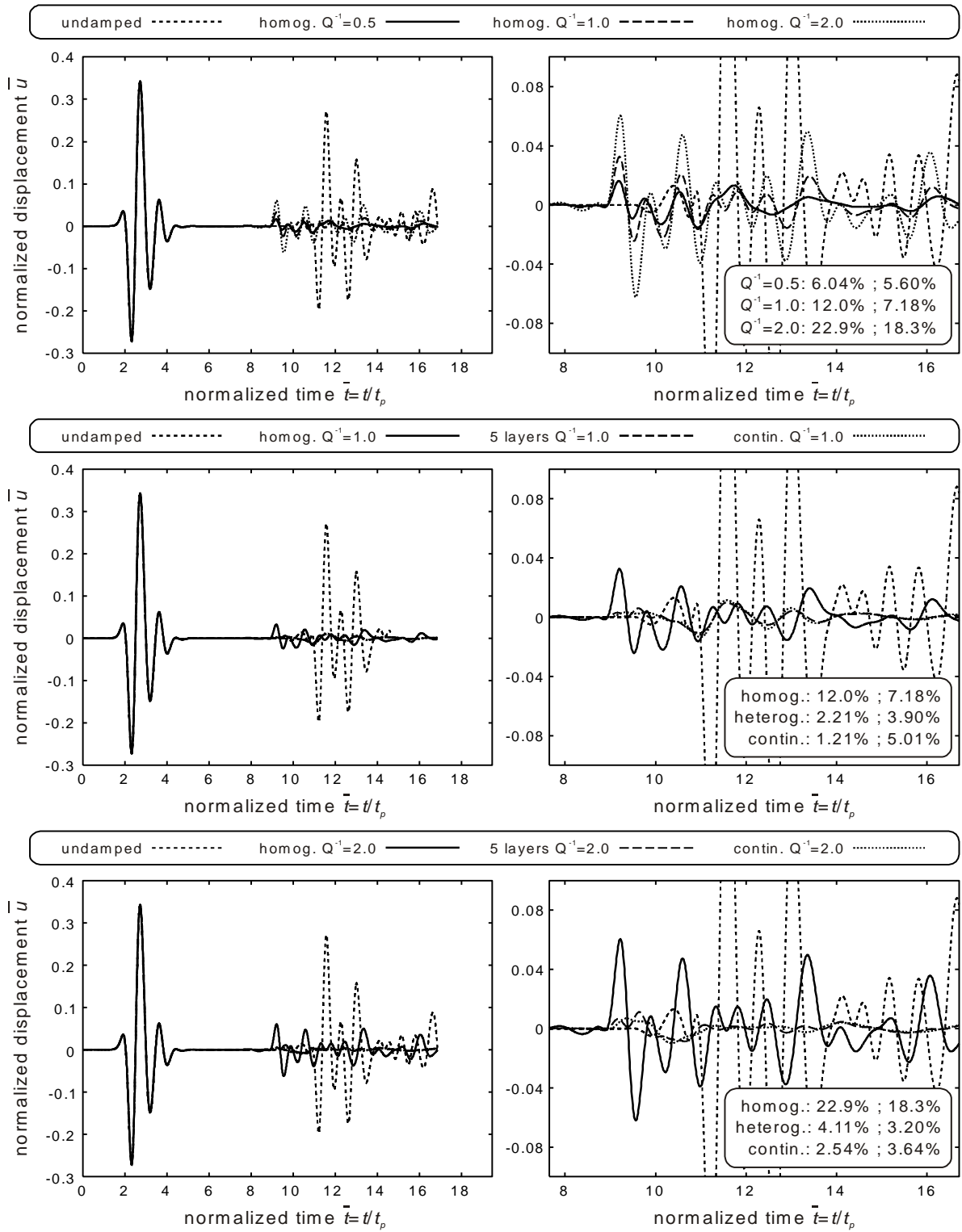

**Fig. 17.** Comparison between homogeneous, heterogeneous and continuous 2D cases at point *A* (surface): curves and maximum relative amplitude of the reflected waves.

In future works, it will be necessary to assess the efficiency of the *Caughey Absorbing Layer Method* in 3D realistic cases and to make some detailed comparisons with other existing methods (e.g. *PML*s). It may be also useful to consider higher order Caughey damping (Eq. (5)) leading to various types of damping-frequency variations. In addition to the Finite Element Method, the *CALM* may be considered in the framework of other numerical methods such as the Spectral Element Method [7,8].